\documentclass[reprint,
 amsmath,amssymb,
 aps, pre
]{revtex4-2}

\usepackage{graphicx}
\usepackage{dcolumn}
\usepackage{bm}
\usepackage{color}

\DeclareMathOperator\erfc{erfc}

\begin{document}

\title{Conditions of structural transition for collisionless electrostatic shock}

\author{Minh Nhat Ly}
 \email{minh-ly@ile.osaka-u.ac.jp}
\author{Takayoshi Sano}
\author{Youichi Sakawa}
\author{Yasuhiko Sentoku}%
\affiliation{%
 Institute of Laser Engineering, Osaka University, Suita, Osaka 565-0871, Japan.
}%

\date{\today}

\begin{abstract}
Collisionless shock acceleration, which transfers localized particle energies to non-thermal energetic particles via electromagnetic potential, is ubiquitous in space plasma. 
We investigate dynamics of collisionless electrostatic shocks that appear at interface of two plasma slabs with different pressures using one-dimensional particle-in-cell (PIC) simulations and find that the shock structure transforms to a double-layer structure at the high density gradient. 
The threshold condition of the structure transformation is identified as density ratio of the two plasma slabs $\Gamma$ $\sim 40$ regardless of the temperature ratio between them.
We then update the collisionless shock model that takes into account density expansion effects caused by a rarefaction wave to improve the prediction of the critical Mach numbers. 
The new critical Mach numbers are benchmarked by PIC simulations for a wide range of $\Gamma$. 
Furthermore, we introduce a semi-analytical approach to forecast the shock velocity just from the initial conditions based on a new concept of the accelerated fraction $\alpha$.
\end{abstract}

\maketitle

\section{Introduction}

The realization of collisionless shock fundamentally changed the field of shock and plasma physics. The collective excitation of plasmas based on wave-particle interactions forms shock structures as a mediating mechanism that replaces collisions and has important implications for the kinetic theory of plasmas \cite{Marcowith2016, Huntington2015}.
On the application side, most astrophysical shocks, from bow shocks at Earth's magnetosphere to supernova remnant shocks \cite{Perri2022, Oka2017, Koyama1995}, are collisionless and thought to be closely related to the origin of cosmic rays \cite{Blandford1987, Blasi2013}. Recent advances in ultra-intense lasers have brought electrostatic shock, a type of collisionless shock, to the laboratory, enabling the generation of high-energy ion beams with a narrow energy band for medical applications \cite{Fiuza2012, Fiuza2013, Haberberger2012}. Because of these attractive applications and the importance of exploring fundamental physics, collisionless shocks have become a topic of increasing interest in recent years \cite{Russell2021, Fiuza2012, Malkov2016, Sakawa2021}.

Studies of shock formation often use particle-in-cell (PIC) simulations to investigate the associated kinetic processes in detail ~\cite{Sorasio2006, Sarri2011a, Fiuza2012, Dieckmann2013b, Russell2021}. 
The shocks can be triggered at interface of two plasma slabs with different pressures in various configurations such as initial density ratios and initial temperature differences. 
The subsequent nonlinear evolution generates a steady-state shock structure characterized by an accompanying electrostatic potential.
Although conditions of the formation of collisionless shocks have been studied extensively, yet an important detail had been overlooked: the emergence of a shock-like structure known as a double-layer instead of a conventional shock structure~\cite{Dieckmann2013}.

In the context of space and astrophysical plasmas, double-layer structures were observed in shock-forming areas such as the Earth's magnetosphere~\cite{Ergun2009, Sun2022} and solar flares~\cite{Li2012} highlighting the need to differentiate these two phenomena.
In general, a collisionless shock or shock structure assumes that particles will move from a lower pressure region (upstream) to a higher pressure region (downstream) in the rest frame of the shock interface. 
In contrast, the opposite flow direction appears in the double-layer structures as ions move from the higher pressure side to the lower pressure side~\cite{Hershkowitz1981b}. By tracking the origin of particles moving from the transition region to the upstream, we can identify the two structures and reconsider the theoretical framework for double-layer structure formation.

The theoretical analysis of electrostatic shocks is usually discussed within Sagdeev's analogy of a particle moving inside a pseudo-potential~\cite{Sagdeev1966}.
Following the framework, we can derive the critical Mach number as the largest possible value for how fast shocks can move.
Over the years, it has been generalized by Sorasio et al.~\cite{Sorasio2006} for non-relativistic temperatures and relativistic temperatures~\cite{Fiuza2012, Fiuza2013, Stockem2013}. 
However, the relationship between the critical Mach number and the Mach number obtained in the simulation has an inconsistency.
The shock speed in simulations exceeds the critical Mach number theoretically predicted in some cases. 
To overcome this inconsistency, we propose a new collisionless shock model that incorporates the previously reported density drop due to plasma expansion~\cite{Sarri2011a, Sarri2011b, Russell2021}. 
This aspect has yet to be analytically included in the current shock model.
In addition, we also present modifications to the mass conservation equation with accelerated fraction as a new parameter to match the ion reflections and the double-layer framework.

The outline of this paper is as follows.
In Sec.~\ref{sec:sec2}, we present PIC simulations including the setup and how we use the results to identify the double-layer structure.
The transition from a collisionless shock to a double-layer structure is found to occur when increasing the initial density ratio. 
The critical value of the density ratio for the transition is also derived.
In Sec.~\ref{sec:sec3}, we introduce the current analytical model and propose the necessary changes to get a more consistent and accurate model. 
First, we consider the density drop due to rarefaction expansion and propose a new model for the critical Mach number.
Then, we obtain predictions of Mach numbers by modifying the mass conservation law to take ion reflection into account.
Finally, in Sec.~\ref{sec:sec4}, we discuss the relationship between the critical Mach numbers and numerical results and the possible interpretation of why the density ratio $\Gamma$ determines collisionless shock or double-layer structure.

\section{\label{sec:sec2} Particle-in-cell simulation}
\subsection{Numerical setup}

\begin{figure}[tp]
\includegraphics[width=0.45\textwidth]{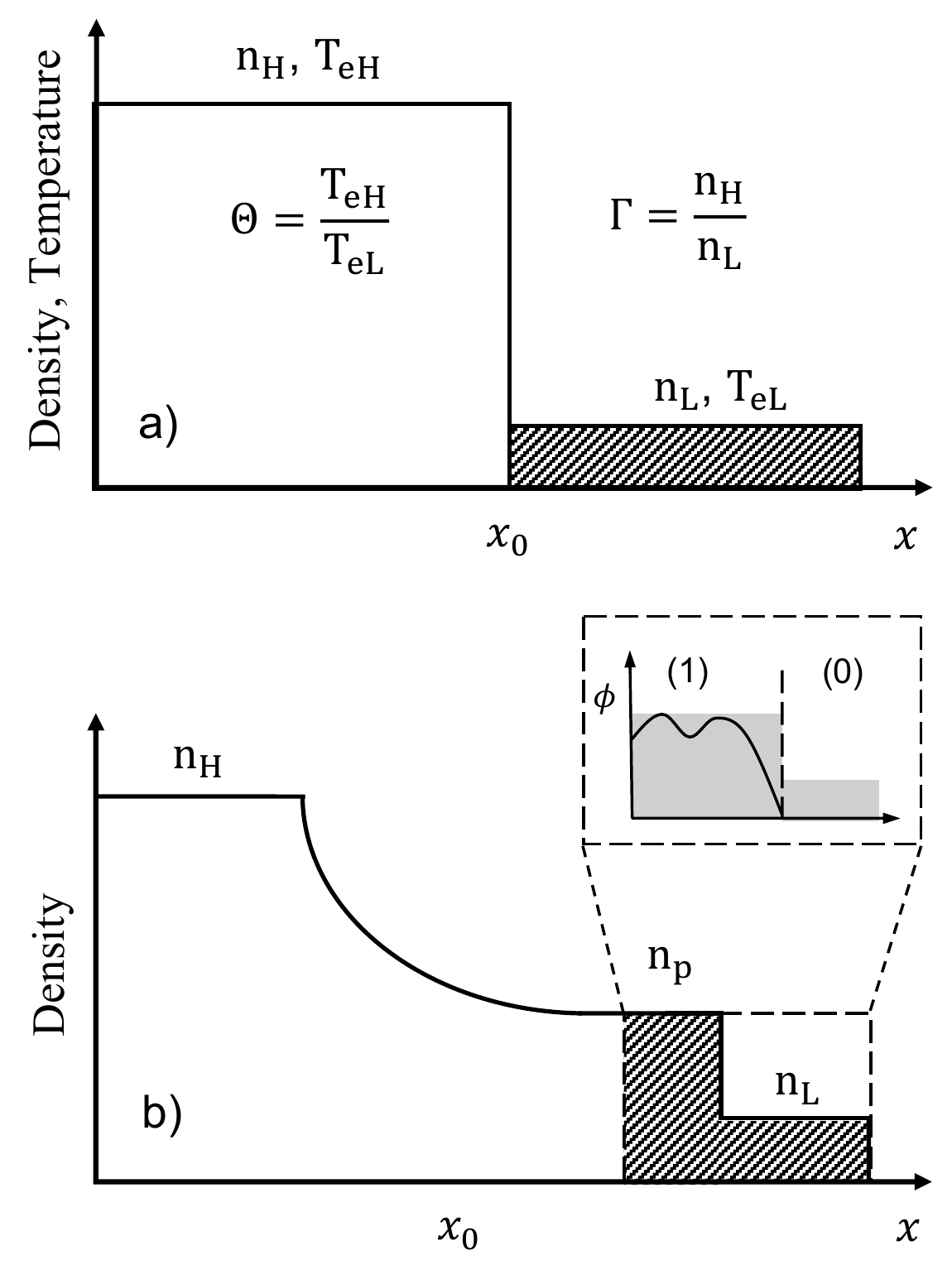}
\caption{\label{fig:shock_structure} (a) The initial setup of PIC simulations is illustrated. Two slabs of plasma with the density ratio $\Gamma$ and temperature ratio $\Theta$ are presented in different shades. (b) A typical ion density profile at the shock's steady-state indicates density reduction near the shock front due to the expansion of the high density slab. The electrostatic potential $\phi$ created by the interactions of two slabs at the front is depicted in the inset. The shock upstream region (low density) and the downstream region are denoted by (0) and (1), respectively.}
\end{figure}

We study shock formation from the interactions of two semi-infinite plasmas by 1D PIC simulations.
The two slabs of plasmas are initiated with different densities and temperatures indicated by the density ratios $\Gamma = n_H/n_L$, and temperature ratios $\Theta=T_{eH}/T_{eL}$.
The schematic figure of our simulation setup is illustrated in Fig.~\ref{fig:shock_structure}(a). Figure~\ref{fig:shock_structure}(b) depicts a typical structure in steady-state showing the expansion front, and shock transition region (downstream and upstream regions in shaded area).
Simulations were performed using a 1D PIC code, PICLS~\cite{Sentoku2008}, with a realistic proton-to-electron mass ratio 1836.
The length of the simulation box is 4000~$c/\omega_{pe}$ with 80000 grid cells and 35-100 particles per cell for each electron and proton.
The simulations ran for a total time of 5000~$\omega_{pe}^{-1}$ with a temporal resolution of $dt = 0.05$~$\omega_{pe}^{-1}$.
The length of the simulation box and running time ensure that electrons from the interface do not yet reach the boundaries during entire calculation.
The reflective boundary condition for particles is applied for simplicity.
In the simulations, the left plasma slab is used as the reference, with its temperature and density remaining constant. While the right slab's density and temperature are varied to change the initial ratios.
The left plasma was initialized with a non-relativistic temperature of $T_{eH} = 10$~keV. 
We start with cold hydrogen ions ($T_i = 0$) for both plasmas.
Two slabs are divided evenly in the computation domain so that the initial interface is located at $x_0=2000\,c/\omega_{pe}$.
In addition, to observe the structural transition of the shocks, we placed particular indexes on particles to label their origin and tracked them in simulations.

As shown in Fig.~\ref{fig:shock_structure}(b), the expansion of plasma causes density reduction from the initial density $n_H$ to $n_p$ in the plateau region near the shock front~\cite{Sarri2011b, Moreno2020}.
Particles in the right slab with density $n_L$ interact with the expanding left slab and form the downstream shock region (1). 
Strictly speaking, the plateau density indicates the density of particles from the left plasma while the downstream region is filled with the right slab's particles. 
Because the two densities have equal values as indicated later in our study [Fig.~\ref{fig:phasespace}(a) and~\ref{fig:phasespace}(d)], we will use the notation $n_p$ for the downstream density in the following discussions.

\subsection{The transition of shock structures}

\begin{figure*}[ht]
\includegraphics[width=0.9\textwidth]{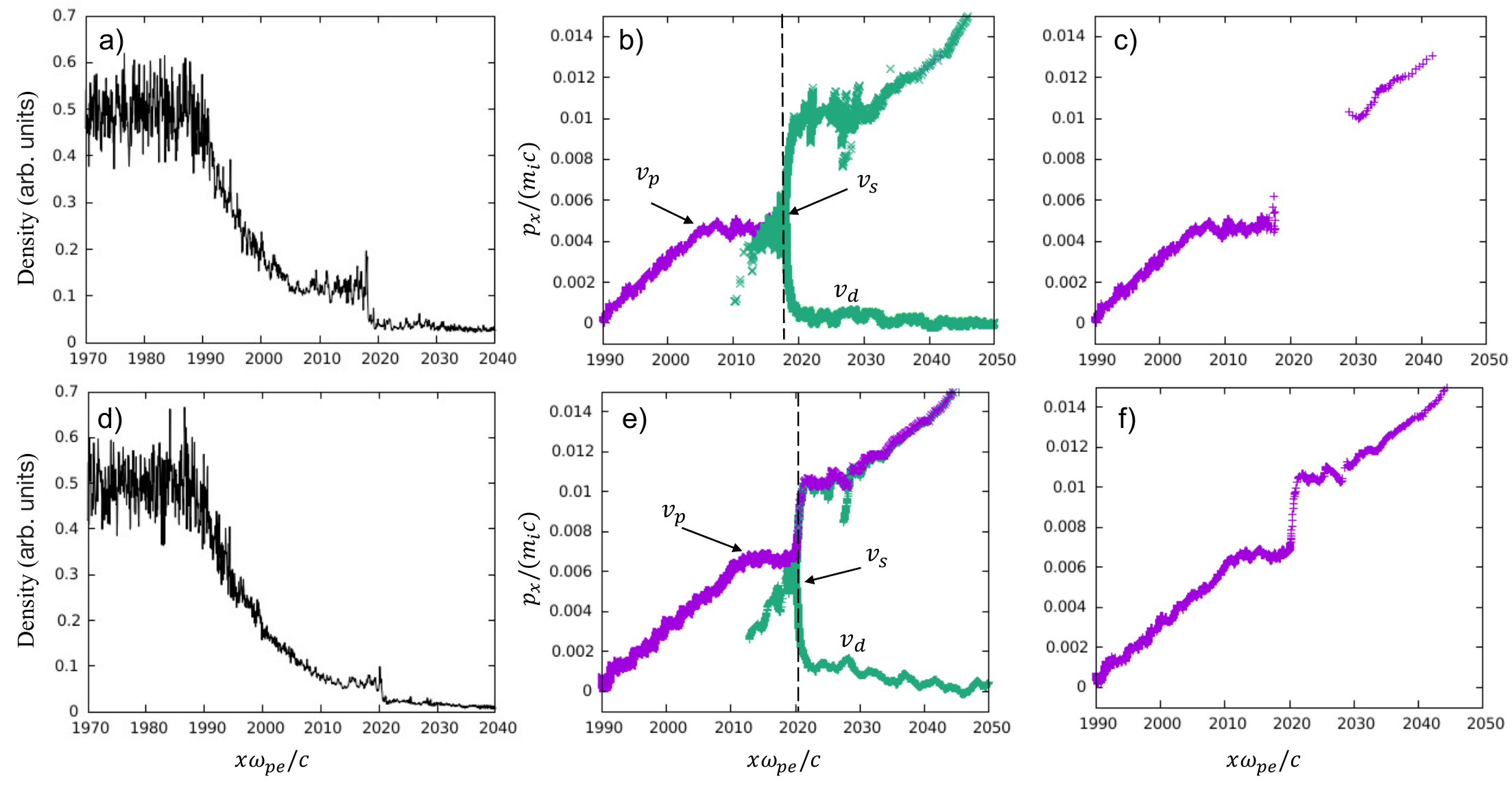}
\caption{\label{fig:phasespace} Ion density profiles and phase space structures from simulations at $t = 3500$ $\omega_{pe}^{-1}$ with $\Theta =1$ ($T_{eH} = 10$ keV), $\Gamma = 20$ for (a) and (b), $\Gamma =100$ for (d) and (e). Ions from the high density slab are represented as purple in (b) and (e), while those from the low density slab are marked as green. (c) and (f) are $p_x$-$x$ diagrams the same as (b) and (e), but only show particles originating from the higher density region.}
\end{figure*}

From the initial setup, the two plasma slabs are released to freely evolve.
The higher density slab on the left expands to the lower density side and a collisionless shock is formed in the transition region [see Fig.~\ref{fig:phasespace}(a) and~\ref{fig:phasespace}(b)].
The shock is electrostatic in nature due to the sufficient temperature and mass ratio between ions and electrons used in our simulations~\cite{StockemNovo2015}.
The shock Mach number, which is the ratio of shock propagating velocity to upstream (lower density region) ion sound speed $M_s = v_s/c_{s0}$, characterizes shock-particle interactions.
In the shock rest frame, ions from the upstream region move toward the shock with the shock velocity and they are decelerated by the electrostatic potential before going to the downstream region.
At the electrostatic shock front, a fraction of ions is reflected back to the upstream region with twice the shock velocity.

Figure~\ref{fig:phasespace} shows the ion density profile and the phase space structures of two simulations (at $t=3500$ $\omega_{pe}^{-1}$) with two different density ratios $\Gamma = 20$ (the upper panels) and $\Gamma = 100$ (the lower panels). A uniform temperature $\Theta = 1$ is assumed for both cases.
The shock velocities $v_s$ are derived from tracking the front position over time.
The shock Mach numbers are $1.60$ and $1.75$ for $\Gamma = 20$ and $100$, respectively.
They are consistent with non-relativistic results reported previously~\cite{Fiuza2013}.
However, when taking the origin of ions into account, these two results exhibit quite different characteristics.
For the case of $\Gamma = 20$, a large fraction of ions from the right slab passes through the shock front and form the shock downstream in the region of $x\sim 2015~c/\omega_{pe}$ as seen in Fig.\,2(b).
The particles from the left slab are not able to reach the shock front just like the fluid shock cases, except for some ions moving before the shock formation [Fig.~\ref{fig:phasespace}(c)], since the expansion plateau velocity $v_p$ is a bit slower than the shock velocity $v_s$.
On the other hand, when $\Gamma = 100$, the particle flow is opposite to the standard shock picture.
Almost all the particles from the right are reflected at the front after the formation of the steady structure.
Instead, as shown in Figs.~\ref{fig:phasespace}(e) and (f), particles from the left overtake the shock front and jump in momentum space, driven by the electrostatic potential.
Similar to the case of the plateau density $n_p$, the plateau velocity $v_p$ is also used to indicate the downstream velocity of the shock. 
Ions in the upstream region ahead of the shock gain net drifting velocity $v_d$ because of the stream of reflected ions [see Fig.~\ref{fig:phasespace}(e)].

\begin{figure}[tp]
\includegraphics[width=0.45\textwidth]{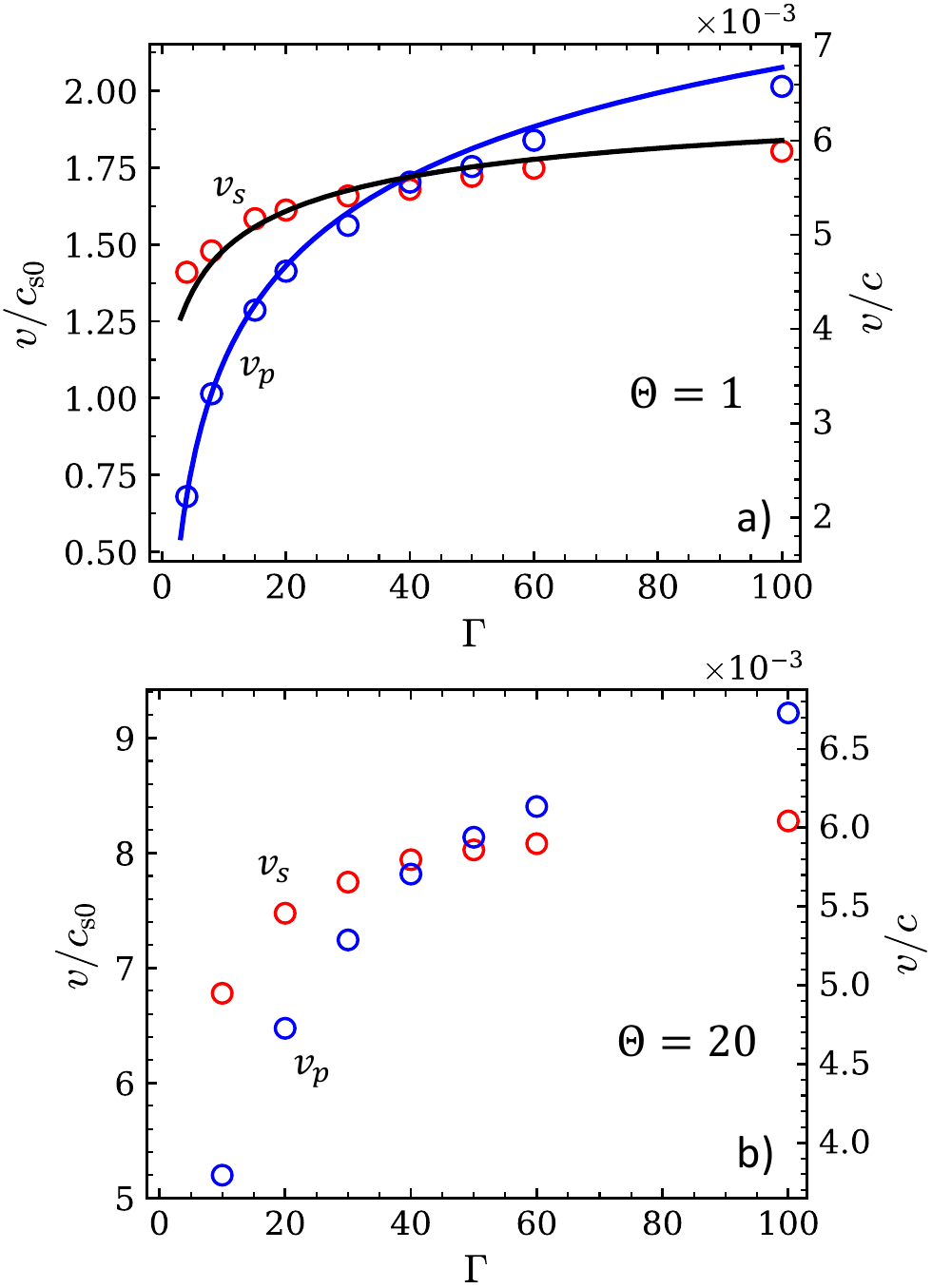}
\caption{\label{fig:Vp_Vs_lab} Comparison between $v_p$ (blue circles) and $v_s$ (red circle) from the simulations (in laboratory frame) for varying $\Gamma = 4 - 100$ with $\Theta =1$ (a) and $\Theta = 20$ (b). The data is illustrated in two scales. The left label is normalized by the sound speed of the upstream plasma $c_{s0}$, which varies with $\Gamma$. The right label is normalized by the speed of light $c$ which are common scale for both panels. In (a), the blue solid line shows $v_p$ from Eq.~(\ref{eq:rfiso2})~\cite{Allen1970, Perego2013} and $v_s$ (black solid line) is calculated from the semi-analytical model discussing in Eq.~(\ref{eq:mass_cons_cs}). It can be seen that the transition occurs around $\Gamma = 40$ in both cases.}
\end{figure}

The opposite flow observed in the case of $\Gamma = 100$ occurs because the expanding velocity of the left slab $v_p$ becomes faster than the shock velocity $v_s$.
As a result, ions from the high density side overtake the shock front and enter the low density region.
To find the critical value $\Gamma$ for such transition, we conducted a series of simulations with varying $\Gamma$ from 4 to 100 for two temperature ratios $\Theta=1$ and $\Theta = 20$. 
The values of $v_s$ ($v_p$) observed in the simulations are plotted by red (blue) circles in Fig.~\ref{fig:Vp_Vs_lab}(a) for $\Theta = 1$ and Fig.~\ref{fig:Vp_Vs_lab}(b) for $\Theta = 20$.
The blue solid line and the black solid line in Fig.~\ref{fig:Vp_Vs_lab}(a) indicate the model for $v_p$ and $v_s$ which will be explored in depth in Sec.~\ref{sec:sec3} B and D, respectively.
It is evident from Fig.~\ref{fig:Vp_Vs_lab}(a) that the plateau velocity $v_p$ exceeds the shock velocity $v_s$ with $\Gamma$ greater than about 40, which is the critical $\Gamma$ initiating the transition from the well-known collisionless shock structure to the other one.
Simulations with different initial temperature ($\Theta = 20$) [Fig.~\ref{fig:Vp_Vs_lab}(b)] yielded no significant impact on the transition and its critical $\Gamma$ which remains around 40 similar to the case with $\Theta = 1$.
Notice that the shock Mach number becomes larger for larger temperature ratios (for $\Theta = 20$ compared to $\Theta = 1$), but the shock velocity has little change.

\begin{figure}[bp]
\includegraphics[width=0.45\textwidth]{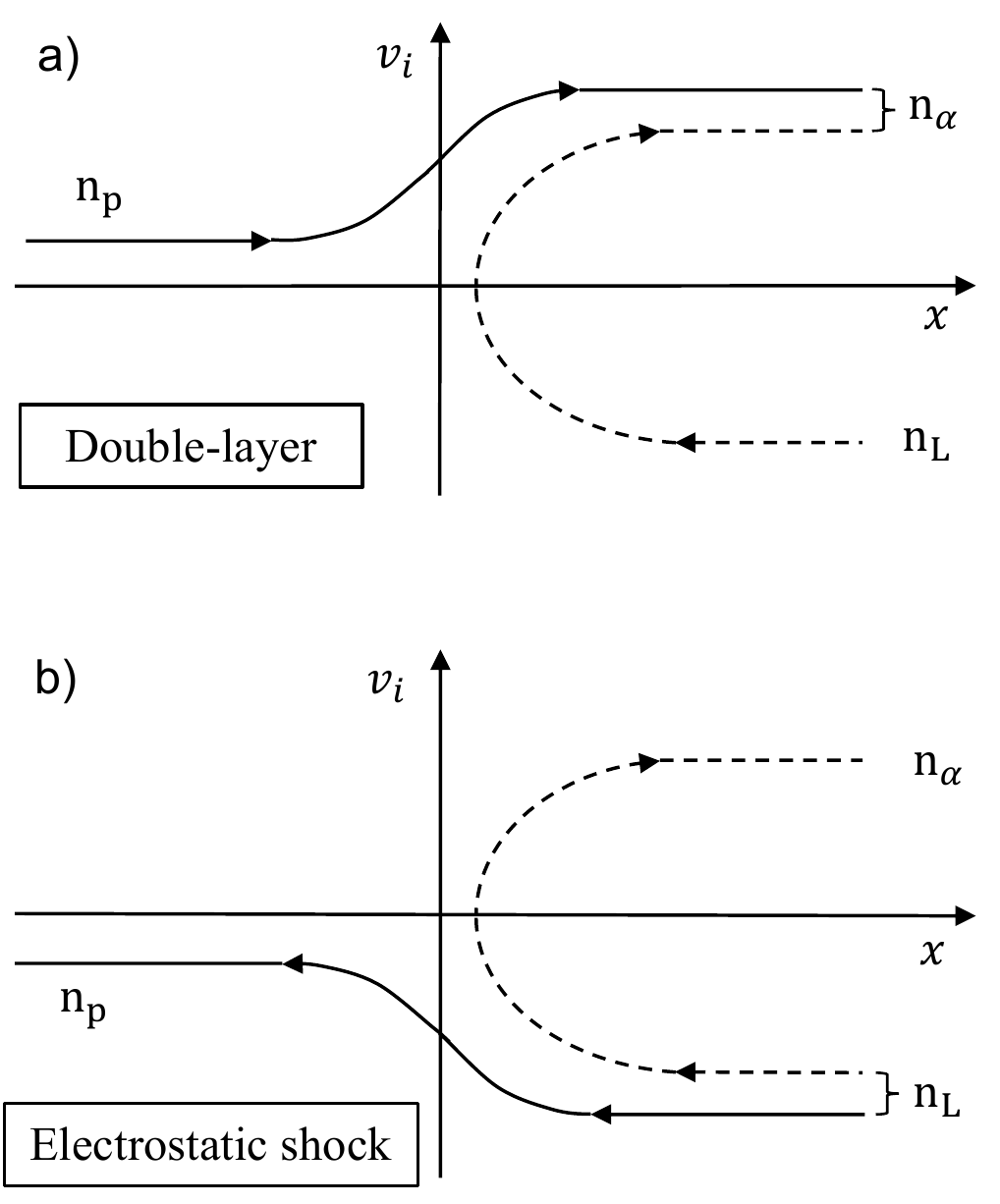}
\caption{\label{fig:dl_cs} Ion phase space in the moving shock frame for double-layer (a) and collisionless electrostatic shock (b) according to Ref.~\cite{Hershkowitz1981b}. In  Sec.~\ref{sec:sec3}.~D, the density of the accelerated ions $n_\alpha$ or ions moving outward from the front, and the accelerated fraction $\alpha = n_\alpha/n_L$ are explained in detail. For electrostatic shocks, $n_\alpha$ is primarily the density of reflected ions while for double-layer structures it also includes the transmitted fraction originating from the high density plasma.}
\end{figure}

Studies of the Earth's auroras and early laboratory experiments identified the structures with the opposite flow as double-layer structures~\cite{Goertz1979, Hershkowitz1981a}.
The conditions to differentiate double-layer and electrostatic shock are given by~\cite{Hershkowitz1981b} with the main difference being the direction of ion flowing through the electric potential called the free ion flow.
Figures~\ref{fig:dl_cs} indicate the schematic phase space structures of double-layer (a) and electrostatic shock (b), in turn, consistent with our simulations with $\Gamma > 40$ and $\Gamma < 40$. 
The electric potential exists in the transition region in both cases but plays different roles in the free ions flow.
In the rest frame of the potential, a double-layer structure is characterized by free ion flow from the higher potential side while the flow from the lower potential side corresponds to the collisionless electrostatic shock.

In the next section, we will discuss how we can modify the conventional theory of collisionless shock to accommodate the structural transition.

\section{\label{sec:sec3}Analytical model for steady-state of electrostatic shocks}

For collisionless electrostatic shocks, analytical models had been developed based on Sagdeev's paradigm~\cite{Sagdeev1966}.
Later studies had explored the role of initial conditions, such as density ratio $\Gamma$ and temperature ratio $\Theta$, on the shock formation~\cite{Sorasio2006, Stockem2013}.
We here focus on non-relativistic shocks and follow closely the approach laid out in Sorasio et al.~\cite{Sorasio2006} which was later adopted in \cite{Fiuza2012, Fiuza2013}.
\subsection{The conventional model for the critical Mach number}
The electric potential of the shocks [the inset from Fig.~\ref{fig:shock_structure}(b)] can be expressed by the one-dimension Poisson equation,
\begin{equation}
 \epsilon \frac{d^2\phi}{dx^2}= e(n_{e} - n_{i})\;.
 \label{eq:Poisson}
\end{equation}
where $\phi(x)$ is the electrostatic potential of the shock, $\epsilon$ is the permittivity of plasma, and $n_i$ ($n_e$) are the ion (electron) density.
To obtain the Sagdeev's characteristic equation, we integrate Eq.~(\ref{eq:Poisson}) with respect to $\phi$ in a condition of $\phi(0) = 0$. The result is the following equation:
\begin{equation}
 \frac{1}{2} \left(\frac{d\varphi}{d\chi} \right)^2 + \Psi(\varphi) =0\;,
 \label{eq:Sageq}
\end{equation}
where we use the normalized units for the potential $\varphi =e \phi/T_{e0}$ with $T_{e0}$ is the upstream electron temperature, and 
 space $\chi = x/\lambda_{D}$ with the Debye length $\lambda_{D} = \sqrt{\epsilon T_{e0}/e^2 n_L}$.
$\Psi(\varphi)$ is the non-linear Sagdeev's potential given by $\Psi(\varphi) = P_i(\varphi) - P_e(\varphi)$ as the difference of ion pressure $P_i(\varphi)$ and electron pressure $P_e(\varphi)$. 

Only when $\Psi(\varphi) < 0$ a shock is developed, and its characteristics, such as its Mach number $M_s$, can then be resolved~\cite{Tidman1971}.
In addition, the shock formation requires the conditions; the potential energy should not exceed the kinetic energy of the shock, $\varphi < M_s^2/2$, otherwise all ions would be reflected. 
With these two conditions the upper limit of the shock Mach number often called the critical Mach number, $M_{\rm cr}$, can be derived. 
The numerical solutions can be obtained by solving $\Psi(\varphi_{\rm max} = M_{\rm cr}^2/2) = 0$ or if we write in full detail~\cite{Fiuza2013}

\begin{widetext}
\begin{equation}
M_{\rm cr}^2 = \frac{1}
{1+\Gamma}
\left\{
\frac{\sqrt{2} M_{\rm cr}}{\sqrt{\pi}}
+ e^{\frac{M_{\rm cr}^2}{2}}\erfc\left(\frac{M_{\rm cr}}{\sqrt{2}}\right)
- 1 + \Gamma\Theta \left[
\frac{\sqrt{2} M_{\rm cr}}{\sqrt{\pi\Theta}}
+ e^{\frac{M_{\rm cr}^2}{2\Theta}}\erfc\left(\frac{M_{\rm cr}}{\sqrt{2\Theta}}\right)
+ \frac{4M_{\rm cr}^3}{3\sqrt{2\pi\Theta^3}} - 1
\right]
\right\}
\;. 
 \label{eq:fiuza13}
\end{equation}
\end{widetext}
where $\erfc$ is the complementary error function.

To accurately derive the shock Mach number $M_s$ from simulation results in the laboratory frame, it is necessary to account for the upstream drift $v_d$. This is because the shock Mach number is defined in the upstream rest frame where the upstream ions are stationary.
To accomplish this, we calculate the shock Mach number as $M_s = (v_s - v_d)/c_{s0}$, where $v_s$ is the shock velocity, and $v_d$ is the upstream drift velocity both measured in the laboratory frame. 
Throughout the rest of the paper, when we mentioned the Mach number $M_s$ without specifying the frame, we have already excluded the drift velocity $v_d$ from the results. 
The results of critical Mach numbers calculated from Eq.~(\ref{eq:fiuza13}) are shown with a black dashed line in Fig.~\ref{fig:new_model} in comparison with our simulation results.

The Mach numbers observed in the simulations (red dots) exceed the critical Mach number given by Eq.~(\ref{eq:fiuza13}) for the cases of $\Gamma > 15$.
This fact indicates that the theoretical model needs to be improved. 
For example, in the previous work, all ions are assumed to pass downstream without reflection. 
Then the partial reflection is not considered in the model. The density expansion effect is also ignored. 
Including these effects is the primary focus of our investigation and will be addressed in detail in the following discussion.

\subsection{The density ratios after the isothermal expansion}
\begin{figure}[bp!]
\includegraphics[width=0.45\textwidth]{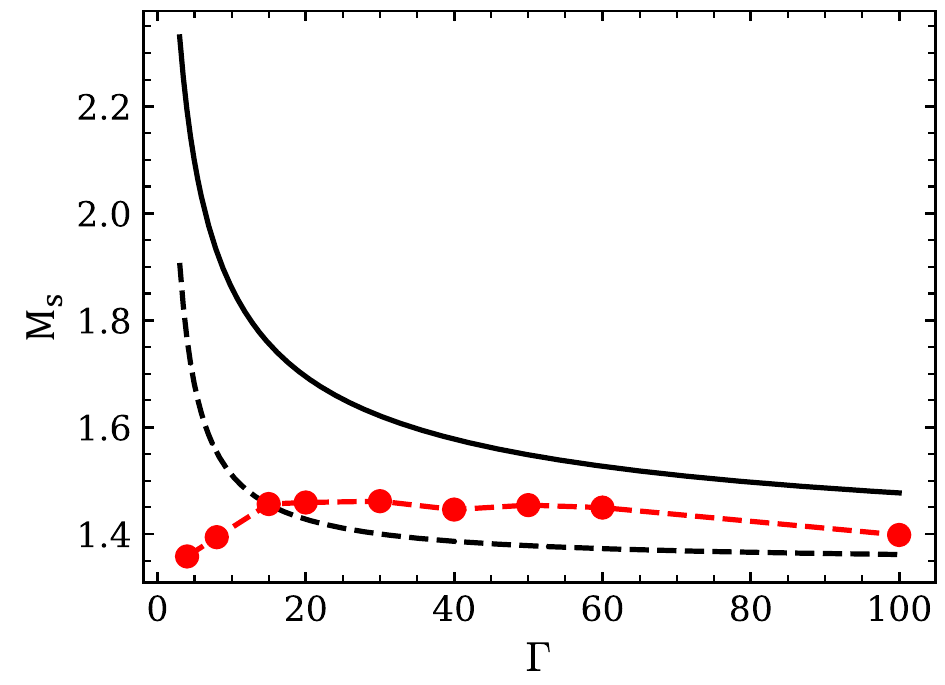}
\caption{\label{fig:new_model}The conventional critical Mach number (black dashed line) is obtained from solving Eq.~(\ref{eq:fiuza13}) with varying $\Gamma$ and $\Theta = 1$. The new critical Mach numbers (black solid line) are calculated using $\Gamma'= n_p/n_L$ calculated from Eqs.~(\ref{eq:rfiso1}) and (\ref{eq:rfiso2}). Simulation results of the shock Mach number $M_s$ with corresponding $\Gamma$ ($\Theta=1$) are shown by red dots. $M_s$ from the simulations exceeds the conventional upper limit but is less than the new model's limit.}
\end{figure}

\begin{figure}[t]
\includegraphics[width=0.45\textwidth]{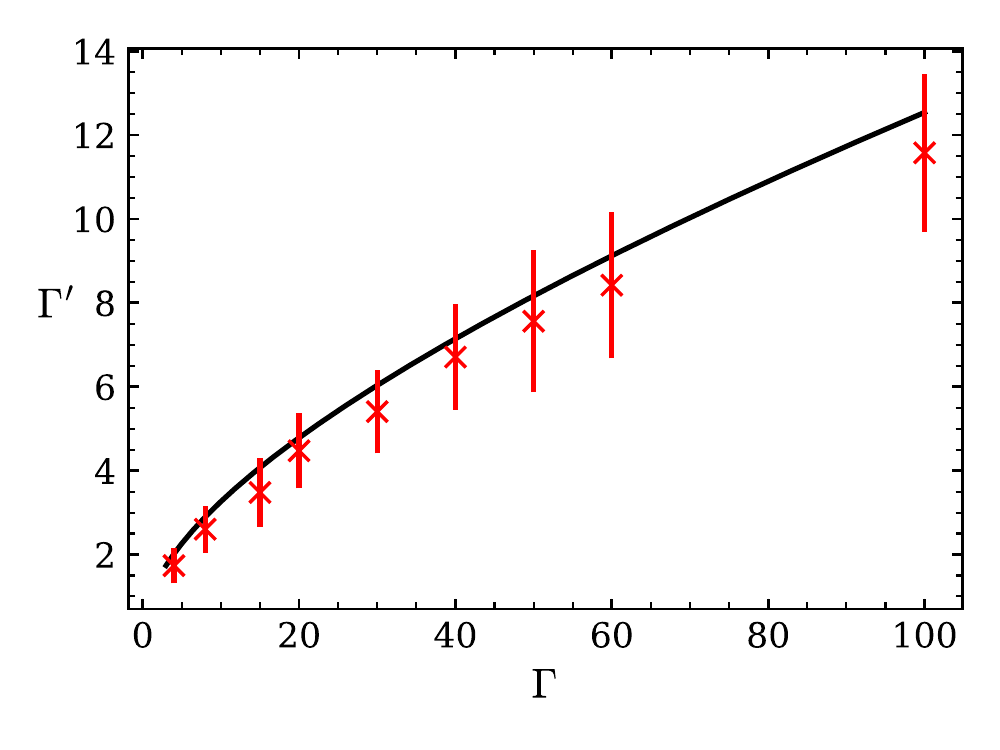}
\caption{\label{fig:expanddens} The density ratios at the expansion front $\Gamma'$ (black solid line) with respect to the initial values calculated by Eqs.~(\ref{eq:rfiso1}) and (\ref{eq:rfiso2}) in comparison with simulation results (red crosses). The bars represent the error evaluated from the density fluctuations in the simulations.}
\end{figure}

In this section, we will present a new shock model that can address a change in the density at the expanding front from the initial value.
As the structural transition from collisionless shocks does not depend significantly on the initial temperature ratio $\Theta$ as showing above, we focus the subsequent discussion solely on the case of $\Theta=1$ for simplicity. 
This implies that the initial sound speed $c_s$ remains constant throughout the region ($c_s = c_{s0} = c_{s1}$).
Nevertheless, a general analysis can be easily performed simply by changing the sound speed when $\Theta \neq 1$.
The previous model assumes that the density ratios at steady-state shock fronts are the same as the initial values. 
While this assumption is valid when the initial density ratio is small, for the higher $\Gamma$, we observed the density ratios drop substantially from the initial values as seen in Fig.~\ref{fig:phasespace}(a) and~\ref{fig:phasespace}(d).

A model for planar isothermal rarefaction waves (see Ref.~\cite{Drake2018}) is applicable to obtain the density ratios in the expansion front, $\Gamma'= n_p/n_L$.
The justification for using the isothermal model is based on the fact that a semi-infinite slab of plasma is sufficiently large to supply heat continuously during the expansion. 
The self-similar solution of the model gives the following relation,
\begin{equation}
 \Gamma' = \Gamma \exp\left(-\frac{v_p}{c_s}\right)\;,
 \label{eq:rfiso1}
\end{equation}
where $v_p$ plays the role of the piston velocity of the expansion. 

The validity of the previous model remains for the steady-state expanding front.
This, in turn, allows us to derive more accurate critical Mach numbers by incorporating the conventional model with the front density ratios $\Gamma'$ as given by Eq.~(\ref{eq:rfiso1}). 
In addition, if we can resolve $v_p$ from initial conditions, we will have an initial value problem $M_{cr}(\Gamma, \Theta)$ similar to the previous model. 

From the well-known quasi-neutral model of gas dynamic (see Ref.~\cite{Allen1970} and Ref.~\cite{Perego2013}), $v_p$ is given by
\begin{equation}
\begin{aligned}
 \left [1-\frac{1}{\Gamma} \exp(v_p/c_s) \right]
 \left [
 \left (\frac{v_p}{c_s} \right)^2 - \frac{2 v_p}{c_s} - 2 \log \left (\frac{1}{\Gamma} \right)
 \right] \\
 -2 \left (\frac{v_p}{c_s} \right)^2 = 0\;.
 \label{eq:rfiso2}
 \end{aligned}
\end{equation}
The plateau velocity $v_p$ obtained from Eq.~(\ref{eq:rfiso2}) is plotted in Fig.~\ref{fig:Vp_Vs_lab}(a) (blue solid line) showing a great agreement with the PIC simulation results.
Using $v_p$ and Eq.~(\ref{eq:rfiso1}), we can calculate the expanding density ratio $\Gamma'$ which is shown in Fig.~\ref{fig:expanddens} with a black solid line in comparison with simulations.
With $v_p$ and $\Gamma'$ derived from hydrodynamic models showing consistency with the simulation results, we can conclude that the plasma far downstream exhibits fluid-like behavior instead of kinetic dynamics as the transition region.
The new critical Mach numbers are obtained by solving Eq.~(\ref{eq:fiuza13}) using $\Gamma'$ instead of $\Gamma$.
The calculated results are illustrated in Fig.~\ref{fig:new_model} with a black solid line.
The critical Mach numbers predicted by the new model are higher than the shock Mach numbers $M_s$ obtained from simulations, satisfactorily serving as the upper limit for allowed shock velocities.

\subsection{Prediction of the transition from collisionless shock to double-layer structures}

\begin{figure}[t]
\includegraphics[width=0.45\textwidth]{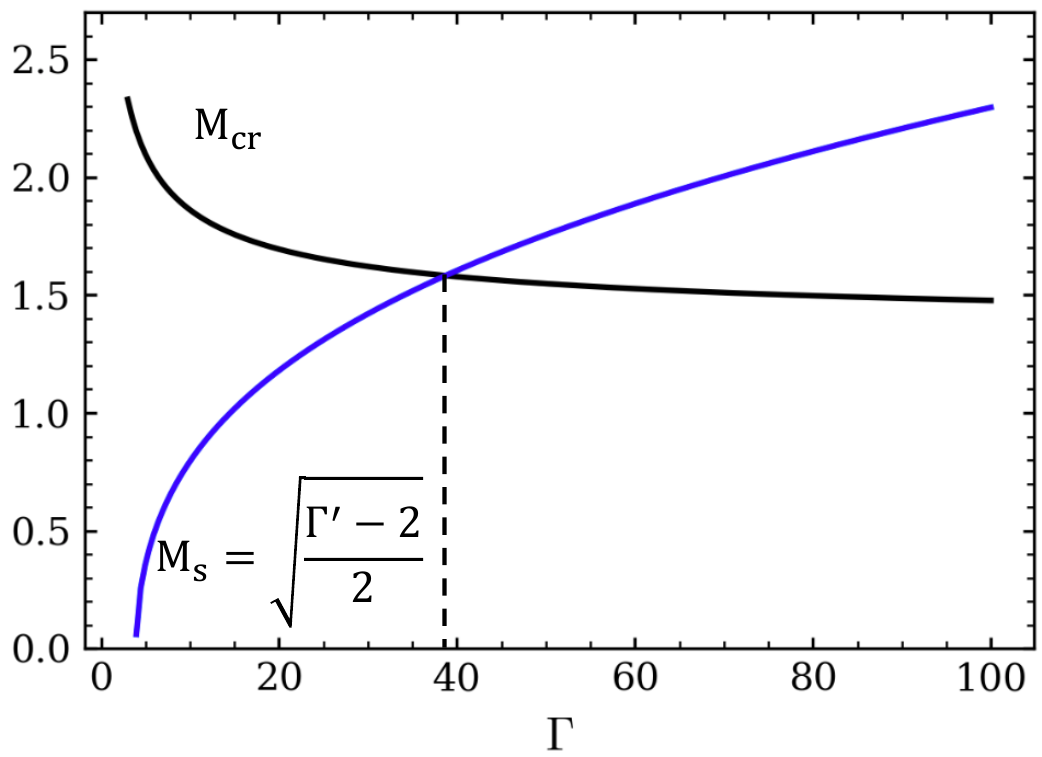}
\caption{\label{fig:transition_gam} The transition condition predicted by Eq.~(\ref{eq:pred_trans_2}) is $\Gamma \sim 40$, which indicated from the crossing point of $M_{\rm cr}$ (black solid line) and $M_s = \sqrt{\Gamma' - 2)/2}$ (blue solid line). $M_{\rm cr}$ is obtained from our new model, and $\Gamma'$ relation with $\Gamma$ is given by the black solid line in Fig.~\ref{fig:expanddens}.}
\end{figure}

The new model of $M_{\rm cr}$ shown in Fig.~\ref{fig:new_model} allows us to predict the transition from the collisionless shock to the double-layer structure.
Here, to evaluate the critical value of $\Gamma$ for the transition, we will derive another condition that needs to be satisfied at the transition.
First, consider the momentum conservation for ions across the shock structure starting from the fluid equation of motion in the steady state,
\begin{equation}
    \frac{\partial}{\partial x} \left ( \rho_i v_i^2 + \mathcal{P}_i \right)
    = -e n_i \nabla \phi\;,
    \label{eq:eqm_conserve}
\end{equation}
where $\mathcal{P}_i$ is ion thermal pressure and $\rho_i = m_i n_i$. 
As we assume cold ions, the contribution of ion thermal pressure here is negligible.
The potential term in Eq.~(\ref{eq:eqm_conserve}) can be expressed using Boltzmann's relation $n_e = n_L \exp(e\phi/T_e)$ with a constant $T_e$. 
Then, the momentum conservation can be expressed approximately as
\begin{equation}
    \frac{\partial}{\partial x} \left (\rho_i v_i^2 + n_i T_e\right)
    \approx 0\;.
    \label{eq:pred_trans_1}
\end{equation}

At the transition, we can assume $v_p = v_s$ which means all ions are reflected without going to the downstream region. 
The reflected ions will have the velocity $v_s$ in the shock frame and the same density as the incoming ions, $n_L$.
Then the jump condition for the transition case with the above conditions gives us the following relation
\begin{equation}
    M_s = \sqrt{\frac{\Gamma' - 2}{2}}\;.
    \label{eq:pred_trans_2}
\end{equation}
With the help of Eqs.~(\ref{eq:rfiso1}) and (\ref{eq:rfiso2}), the right-hand side of Eq. (8) is given as a function of $\Gamma$, which is shown in Fig.~\ref{fig:new_model}.
The crossing point of the critical Mach number $M_{\rm cr}$ and $M_s$ indicates the maximum value of $\Gamma$ for existing collisionless shock.
This critical value of $\Gamma \approx 40$ is consistent with our PIC simulation result depicted in Fig.~\ref{fig:Vp_Vs_lab}.

\subsection{Prediction of the Mach numbers}

\begin{figure}[tp!]
\includegraphics[width=0.45\textwidth]{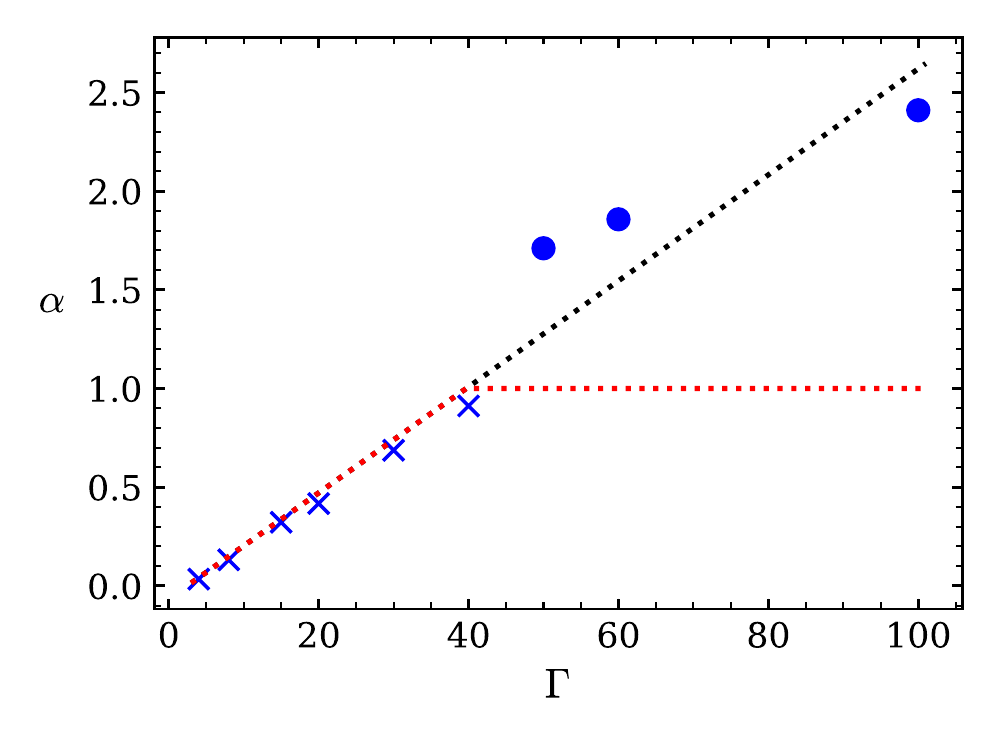}
\caption{\label{fig:shock_mach_pred}accelerated fraction $\alpha=n_{a}/n_L$ obtained from the simulations with respective $\Gamma$ and $\Theta = 1$. We derive $\alpha$ from tracking the reflection rate for collisionless shock cases (blue crosses), while $\alpha$ for double-layer (blue dots) is obtained by the outflow density $n_\alpha$ [see Fig.~\ref{fig:dl_cs}(a)]. Black dot line is the extension of collisionless shocks' accelerated fraction which is used with Eq.~(\ref{eq:mass_cons_cs}) to derive $v_s$ in Fig.~\ref{fig:Vp_Vs_lab}(a). The reflection rate is denoted by the red dot line.}
\end{figure}

In hydrodynamic shocks, if the downstream velocity and the density jump have already known, the shock Mach numbers can be obtained by the mass conservation law of the jumps conditions. However, it is not that straightforward for collisionless shocks and double-layer structures. In order to derive the conservation, factors such as reflected ions in the strong shocks and the reversed flow in double-layer structures need to be taken into account. 

Let's define $\alpha \equiv n_a/n_L$ as the accelerated fraction, where $n_a$ is the density of particles accelerated by the electric field at the expansion front (either collisionless shock or double-layer) to twice the front velocity (Figs.~\ref{fig:dl_cs}).
In the case of collisionless electrostatic shocks, $n_\alpha$ represents the density of reflected ions [dashed line in Fig.~\ref{fig:dl_cs}(b)], and $\alpha$ can be regarded as the reflection rate of incoming ions. Mass conservation applies to the fraction of ions passing to the shock downstream without being reflected $(1-\alpha)n_L$ (solid line in the Fig.~\ref{fig:dl_cs}(b)), hence, can be written as
\begin{equation}
 n_p (v_s - v_p) = (1-\alpha)n_L v_s\;.
 \label{eq:mass_cons_cs}
\end{equation}

In the double-layer regime, $n_\alpha$ is a sum of two components: the reflected ions [dashed line in Fig.~\ref{fig:dl_cs}(a)] and the transmitted ion moving to the low density slab [Fig.~\ref{fig:dl_cs}(b)].
Since almost all upstream ions are reflected, we have $\alpha = (1 + n_T/n_L)$ with $n_T$ as the transmitted ion density. 
As the transmitted flow connects two plasma regions, mass conservation should be derived based on this component.
Thus, we get the relation $n_p (v_p - v_s) = n_T v_s$.
Using the definition $n_T = (\alpha - 1) n_L $, we retrieve the same expression as Eq.~(\ref{eq:mass_cons_cs}).

Ions reflected from the shock with twice its velocity $2 v_s$ are the reason we can have the same relation for both collisionless shocks and double-layer structures.
The validity of Eq.~(\ref{eq:mass_cons_cs}) for double-layers strongly depends on the stability of the upstream structure.
For instance, in large $\Gamma$ cases, the velocity of the reflected and reversed flow fraction increase to exceed $2 v_s$ [Fig.~\ref{fig:phasespace}(e) and (f) for $\Gamma = 100$] so we expect to see discrepancies between prediction values and simulation results.

We can predict the shock Mach numbers from Eq.~(\ref{eq:mass_cons_cs}) from the initial condition (the density ratio $\Gamma$) if we knew the accelerated fraction $\alpha$. 
However, there is no analytical model allowing us to obtain the accelerated fraction $\alpha$ at this moment.
For a semi-analytical prediction of shock Mach numbers, we can use the results of $\alpha$ from PIC simulations which are indicated in Fig.~\ref{fig:shock_mach_pred}. 
The values of $\alpha$ for the collisionless shock regime ($\alpha \leq 1$) are obtained from tracking the reflection ratio and indicated by the blue crosses in Fig.~\ref{fig:shock_mach_pred}. 
For the double-layer regime ($\alpha \geq 1$), the values are illustrated in the blue dots by comparing the transmitted fraction $n_T$ and $n_L$.
Although we can have a unified definition of $\alpha$, we can recognize the results in the two regimes are loosely related and we can see each of them scaled somewhat differently from Fig.~\ref{fig:shock_mach_pred}.
Additionally, it is easy to see that $\alpha$ increases linearly with respect to $\Gamma$.
In this work, we applied the linear fitting of $\alpha$ but only for $ \alpha< 1$ or the shock regime (black solid line in Fig.~\ref{fig:shock_mach_pred}) for later prediction of shock Mach numbers.
The justification is that for $\alpha > 1$ or the double-layer regime, the validity of Eq.~(\ref{eq:mass_cons_cs}) is marginal because of the increase in velocity of accelerated fraction as discussed above.
The predicted shock Mach number shown as the black solid line Fig.~\ref{fig:Vp_Vs_lab}(a) is consistent with PIC simulations (red circles) proving that our approximation for $\alpha$ is indeed a reasonable choice.

\section{\label{sec:sec4}Discussion}

\begin{figure}[bp!]
\includegraphics[width=0.45\textwidth]{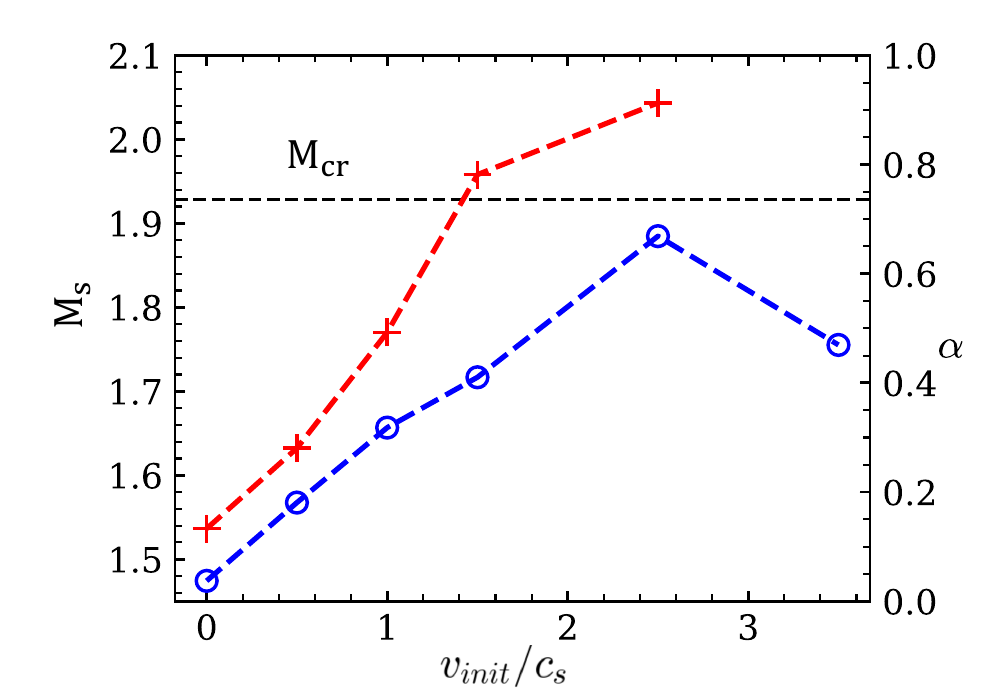}
\caption{\label{fig:Hg8_drift} Shock Mach numbers (blue circles) and accelerated fraction $\alpha$ (red crosses) when adding initial velocity $v_{init}$ for the case of $\Gamma = 8$, $\Theta = 1$ and $T_e = 10$ keV. The critical Mach number (black dashed line) is approximately $1.92$ in this case. We can see that when $M_s$ reach the critical value, $\alpha \approx 1$ as predicted from the critical condition $\varphi = M^2_{cr}/2$ and the definition of $\alpha$. Shock structures start to saturate with $v_{init}> 2.5~c_s$. At $v_{init} = 3.5~c_s$, we have $\alpha \approx 6.3$ and it is no longer a collisionless shock structure.}
\end{figure}

The condition to obtain the critical Mach numbers, $\varphi = M^2_{cr}/2$, implies that the downstream velocity in the shock frame $v_p = v_s$ which also means $\alpha = 1$ or a perfect reflection [Eq.~(\ref{eq:mass_cons_cs})].
We observed that $\alpha$ increases as $v_p$ increases but $v_p$ obtained from two static slabs (without any initial relative velocity between them) is limited by the initial density jump $\Gamma$.
However, previous works showed that for a given $\Gamma$, we can increase $v_p$ and the reflection rate by adding initial colliding drift $v_{\rm init}$ between two plasma slabs~\cite{Sorasio2006, Fiuza2013}.
Fiuza et al.~\cite{Fiuza2013} also define $M_{cr}$ as Mach numbers when ion reflection is observed as a consequence of increasing $v_{\rm init}$.
However, a more consistent model should not only see ion reflection at $M_{cr}$ but a perfect reflection $\alpha \approx 1$ due to the implication of $M_{cr}$ mentioned above.
An example of the argument is indicated in Fig.~\ref{fig:Hg8_drift} for $\Gamma = 8$. It is clear that increasing the initial velocity $v_{\rm init}$ leads to a larger shock Mach number and accelerated fraction $\alpha$ (also reflection fraction in this context).
The trend continues until shock Mach numbers approach the critical value and $\alpha \approx 1$ ($v_{init} = 2.5~c_s$).
For $v_{init} > 2.5~c_s$, the flow becomes too fast for the shock formation and the shock structure saturated into weakly perturbed flows.
We can still numerically derive $\alpha$ by its definition which becomes greater than 1. 
For instance, $\alpha \approx 6.3$ with $v_{init} = 3.5~c_s$ and the structure is no longer collisionless shock in this case.

\begin{figure}[tp!]
\includegraphics[width=0.45\textwidth]{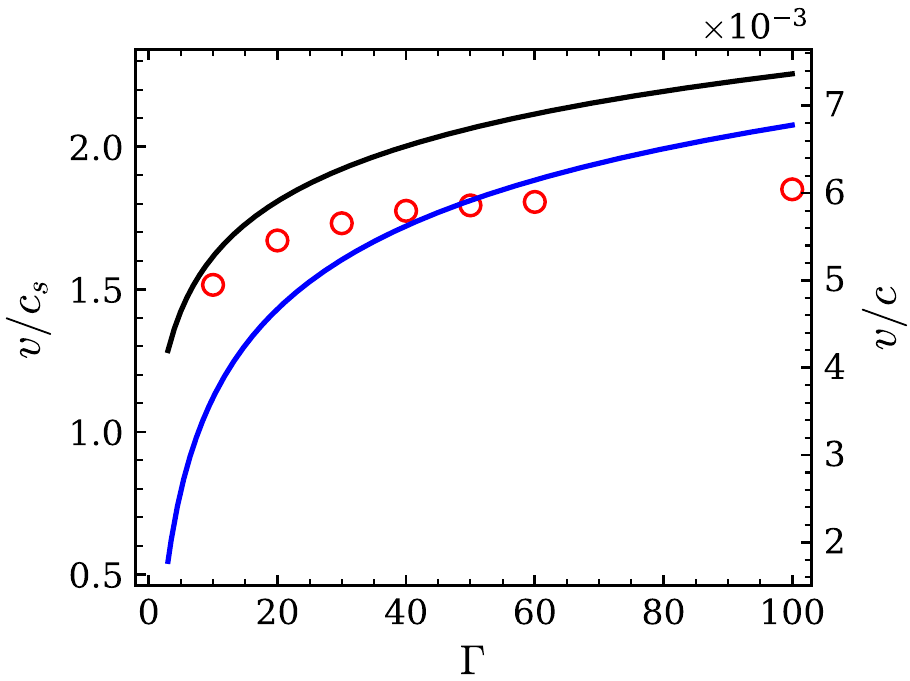}
\caption{\label{fig:hydro_shock_comp} Comparison of the shock velocity between collisionless shock (red crosses) and collisional shock Mach numbers (black solid line) calculated from Eq.~(\ref{eq:mass_cons_cs}) with $\alpha = 0$ (no reflection). As a reference, $v_p$ is also shown by the blue solid line. It is clear that, unlike collisionless shocks, $v_s$ is always greater than $v_p$ for collisional shocks, implying that there is no transition to double-layer structures.}
\end{figure}

Further research is needed to shed insights into the transition from collisionless shock to double-layer. 
For now, let us briefly mention one possible interpretation of the transition based on the macro-view of shocks as a means of dissipating energy for material crossing the shocks. 
From this perspective, the reversed flow could be seen as a mechanism for the shocks to release excessive energy created from the input power (initial density ratios $\Gamma$).
In fact, this argument can be seen from our simulations with increasing $\Gamma$.
For small input power, the shocks only dissipate their energy by accelerating upstream particles to downstream velocity (in laboratory frame).
When the density ratios increase ($\Gamma \sim 10 - 40$) and exceed the limit provided by the previous method, the shock simply rejects the injections of energy by reflecting ions back to the upstream region~\cite{Balogh2013physics}. 
Finally, when all the upstream ions are reflected ($\Gamma \gtrsim 40$), the ions from the high density plasma have to emit their own energy, hence the reversed flow appears.

The reflection of upstream particles is the special feature that distinguishes collisional and collisionless shocks.
Given that the collision rate is always sufficient to dissipate energy for the shock, we can see that increasing the input power simply increases the shock velocity, with no reflections.
To prove this point, we can calculate the collisional shock velocity by using Eq.~(\ref{eq:mass_cons_cs}) with $\alpha = 0$ as there is no reflection ($v_p$ is similar to the values of collisionless shock).
As shown in Fig.~\ref{fig:hydro_shock_comp}, the collisional shock velocity is always larger than $v_p$, implying no transition to double-layer structures.

\section{Summary}
In summary, we have studied the transition of the well-established collisionless shock to the double-layer structure.
By using 1D PIC simulations, we prove that double-layer structures can emerge under the same two-slab plasma configurations which often use to study electrostatic shock formations.
We found the transition occurs around $\Gamma = 40$ and is independent of the initial temperature ratio $\Theta$.
We then discussed the differences between the two structures and the ideas of how such a transition can happen.

In the second part, we proposed a new model for the critical Mach number based on previous works~\cite{Sorasio2006, Fiuza2013}.
The main feature of our model is the incorporation of the density expansion that reduces the actual density ratios at the front.
In particular, we use of the realistic density ratio at the front $\Gamma'$ instead of the initial density ratio $\Gamma$ for a more consistent description of the shock structure.
The new model offers a promising result as it addresses previous discrepancies between the critical Mach number and the Mach number observed in simulations.
In addition, our model can predict the critical value of $\Gamma$ for the transition from collisionless shock to double-layer consistently with the numerical results.

Finally, we introduce a new concept of $\alpha$ with the motivation to include the ion reflection into the shock jump condition.
The concept can also provide a consistent description of the front velocity from collisionless shock to double-layer cases.

\section*{Acknowledgement}
This study was supported by JSPS KAKENHI Grants No. JP19KK0072, and No. JP20H00140. MNL would like to express his gratitude to Okazaki Kaheita Foundation for providing financial support for his study and research.

\bibliography{aps}

\end{document}